\documentstyle[namedreferences,epsfig]{kluwer}

\newcommand{\et}{{\em et al.}}

\begin{opening}

\title{Synchronization and oscillatory dynamics in heterogeneous,
mutually inhibited neurons}

\author{John A. \surname{White}}
\institute{Department of Biomedical Engineering,
Boston University, Boston, MA 02215}
\author{Carson C. \surname{Chow}}
\author{Jason \surname{Ritt}}
\author{Cristina \surname{Soto-Trevi\~no}}
\author{Nancy \surname{Kopell}}
\institute{Department of Mathematics,
Boston University, Boston, MA 02215}

\date{\today}

\end{opening}

\begin{document}

\maketitle

\begin{abstract}
We study some mechanisms responsible for synchronous 
oscillations and loss of synchrony at physiologically relevant
frequencies (10-200 Hz)   
in a network of heterogeneous inhibitory neurons.
We focus on the factors that determine the level of synchrony and
frequency of the network response, as well as the effects of 
mild heterogeneity on network dynamics.  With
mild heterogeneity, synchrony is never perfect and is relatively
fragile. 
In addition, the effects of inhibition are more complex in 
mildly heterogeneous networks than in homogeneous ones.
In the former, synchrony is broken in two distinct 
ways, depending on the ratio of the synaptic decay time to the 
period of repetitive action potentials ($\tau_s/T$), where
$T$ can be determined either from the network or from a single,
self-inhibiting neuron.
With $\tau_s/T > 2$, corresponding to large applied current, small
synaptic strength or large synaptic decay time, the effects of inhibition 
are largely tonic and heterogeneous neurons spike relatively independently.
With $\tau_s/T < 1$, synchrony breaks when faster cells begin to
suppress their less excitable neighbors; cells that fire remain nearly
synchronous.  
We show numerically 
that the behavior of mildly heterogeneous networks can be related to
the behavior of single, self-inhibiting cells,
which can be studied analytically.
\end{abstract}

\section{Introduction}
Synchronous activity has been observed in many regions of the
brain and has been implicated as a correlate of
behavior and cognition~\cite{gray94,llinas93}.  In
the hippocampal formation, where such activity has been
studied most thoroughly, neurons
discharge in several behaviorally important synchronous
rhythms~\cite{buzsaki86}.
Among these patterns are the theta (4-12 Hz)
and gamma ($20-80$ Hz) rhythms, which appear as nested rhythms under
conditions of active exploration and paradoxical sleep, as well as
hippocampal sharp waves ($\sim 0.5$ Hz), which occur along with
embedded fast ripples ($\sim 200$ Hz) under conditions of rest and
slow wave sleep~\cite{bragin95,ylinen95}.  Here, we investigate some
mechanisms responsible for generating synchronous oscillations throughout
the physiologically relevant range of frequencies (10-200 Hz).

Two crucial results point to the importance of inhibitory
interneurons in generating synchronous rhythms in the hippocampal
formation.  First, it has been shown in intact animals that interneurons 
fire robustly and synchronously in both the theta-gamma state and
in the sharp wave-ripple state~\cite{bragin95,ylinen95}.  
Second, {\it in vitro} experiments
have demonstrated that a functional network containing interneurons 
alone can support synchronous gamma activity~\cite{whit95}.
These and other experimental results have spurred both 
analytic~\cite{ernst,hansel95,gerstner96b,vrees94}
and numerical~\cite{jeff96,traub96,traub96b,wang96,whit95}
studies of synchrony among neurons.
Among the principal conclusions of such studies are that
stable synchrony is supported by inhibition that is slow
compared with neuronal firing rates; and that firing rate 
decays linearly, eventually saturating, as a function of the decay 
time constant of inhibition ($\tau_s$). 
When the synaptic coupling is
extremely fast,  the coupling tends to push the
neurons towards
anti-synchrony~\cite{friesen94,perkel74,skinner94,vrees94,wang92}.

Synchronous oscillations generated {\it in vivo} are almost certainly 
the product of interactions among neurons with some (unknown) degree 
of heterogeneity in excitatory drive and intrinsic excitability.
Much of the earlier work in the area
has not explored the effects of heterogeneity in intrinsic spike
rates~\cite{ernst,gerstner96b,jeff96,traub96,traub96b,vrees94,whit95}.
Tsodyks \et (1993) considered a network of 
integrate-and-fire oscillators with heterogeneous external drive and
all-to-all
{\em excitatory} coupling. They found that for an infinite number of
oscillators, those with an external drive below a critical value would be
synchronized and those above the critical value would be asynchronous.
This co-existence around the critical value persisted in the limit of
vanishing heterogeneity. 
Golomb and Rinzel (1993) considered a heterogeneous network of
all-to-all coupled
inhibitory bursting neurons and found regimes of synchronous,
anti-synchronous and asynchronous behavior when the width of the
heterogeneity was changed. 
They considered a parameter regime that was synchronous for small
heterogeneity.  Wang and
Buzs\'aki~(1996) considered a 
hippocampal interneuron network with heterogeneity in the external
drive and network connectivity.  They found numerically that for a
physiologically plausible parameters, coherent activity is only
possible in the gamma range of frequencies.

Our purpose here is to understand more fully the implications of
small levels of heterogeneity 
for the degradation of synchrony in networks of inhibitory fast
spiking neurons 
and the mechanisms by which this degradation occurs.
To this end, we have begun a coordinated set of analytic and numerical
studies of the problem.  In this paper, we numerically analyze a
network of interneurons applicable to the CA1 region of the
hippocampus.  We consider slow inhibition and heterogeneity in the
external drive.  We find that small amounts of
heterogeneity in the external drive can greatly reduce coherence. 
In addition,  we find that coherence can be reduced in two qualitatively
different ways depending on the parameters -- either by a transition to
{\em asynchrony} where the cells fire independently of each other, or
through 
{\em suppression} where faster cells suppress slower cells. 

The reaction of a network to heterogeneity is shown in the paper
to be correlated with the dependence of firing frequency on the
time constant of synaptic decay.
We find in self-inhibiting cells or synchronous networks
that this dependence divides into two asymptotic regimes.  
In the first (the tonic-inhibition or {\em tonic} regime),
inhibition acts as if it were steady-state and only weakly affects discharge
frequency.  
In the second (the phasic-inhibition or {\em phasic} regime), 
time-varying
inhibition firmly controls discharge frequency.
There is a gradual crossover between these regimes.
The presence of
a neuron or network in the tonic or phasic regime can most easily
be determined  
by examining the ratio of the synaptic decay time constant
to discharge period ($\tau_s/T$).
(Discharge period $T$ can be obtained from the full network
or from a reduced model including only a single cell with 
self-inhibition.)
$\tau_s/T$ is large ($> 2$ for our parameters) and 
varies linearly with $\tau_s$ 
in the tonic regime.
$\tau_s/T$ is small ($< 1$) and only logarithmically 
dependent on $\tau_s$ in the phasic regime.  
However, if $\tau_s$ is {\em too} small ($<< 1$), the phasic regime
is departed and anti-synchrony is possible.
Networks of weakly heterogeneous (less than 5\%)  cells
generally exhibit   
asynchrony (defined here as the state of phase dispersion) 
in the tonic regime. 
In the phasic regime, cells generally exhibit a form of locking,
including synchrony, harmonic locking (locking at rational ratios),
and suppression.
These results can be demonstrated
analytically using a reduced model 
with mutual and self-inhibition~\cite{cc}.

We conclude 
that mild heterogeneity in inhibitory networks adds effects
that are not accounted for in previous analyses, but that are
tractable under our current framework.
In particular, we show that the prediction 
that slow inhibition leads to synchrony, 
made under assumptions of 
homogeneity~\cite{ernst,gerstner96b,vrees94},
must be modified in the presence of mild heterogeneity.
Thus, the new framework provides a context for understanding previous 
simulations~\cite{wang96}.
In particular, it explains the mechanisms 
underlying asynchrony (phase dispersion) with slow 
decay of inhibition.
These mechanisms differ from those underlying the loss of synchrony with
faster-decaying inhibition.  

\section{Methods}

\subsection{Numerical simulations}

Simulations were carried out using single-compartment neurons with
inhibitory synapses obeying first-order kinetics.
Membrane potential in each point neuron obeyed the current balance
equation
\begin{equation}
C\frac{dV_i}{dt}=I_i-I_{Na}-I_K-I_L-I_s,
\label{membrane}
\end{equation}
where $C=1 \mu$F/cm$^2$, $I_i$ is the applied current,
$I_{Na}=g_{Na} m_\infty^3 h (V_i - V_{Na})$ and $I_K=g_Kn^4(V_i-V_K)$ are
the Hodgkin-Huxley type spike generating currents, $I_L= g_L(V_i-V_L)$
is the leak current and $I_s = \sum_j^N (g_s/N) s_j (V_i-V_s)$ is the
synaptic current. 
The fixed parameters used were: $g_{Na}=30$~mS/cm$^2$, $g_K=20$~mS/cm$^2$, 
$g_L=0.1$~mS/cm$^2$,
$V_{Na}=45$~mV, $V_K=-75$~mV, $V_L=-60$~mV, $V_s=-75$~mV.
These parameters are within physiological ranges and give the
high spike rates typical of hippocampal interneurons.  
The phenomena described here seem largely independent of specific
neuronal parameters.

The activation variable $m$ was assumed fast and substituted with
its asymptotic value
$m_\infty(v)=(1 + \exp[-0.08(v+26)])^{-1}$.  The inactivation
variable $h$ obeys
\begin{equation}
\frac{dh}{dt}=\frac{h_\infty(v)-h}{\tau_h(v)},
\end{equation}
with $h_\infty(v)=(1 + \exp[0.13(v+38)])^{-1}$, 
$\tau_h(v)=0.6/(1 + \exp[-0.12(v+67)])$.  The variable $n$ obeys
\begin{equation}
\frac{dn}{dt}=\frac{n_\infty(v)-n}{\tau_n(v)},
\end{equation}
with $n_\infty(v)=(1 + \exp[-0.045(v+10)])^{-1}$, 
$\tau_n(v)=0.5+2.0/(1+\exp[0.045(v-50)]))$.

The gating variable $s_j$ for the synapse is assumed to obey first order
kinetics of the form
\begin{equation}
\frac{ds_j}{dt} = F(V_j)(1-s_j)-s_j/\tau_s,
\label{synapse}
\end{equation}
where $F(V_j)=1/(1+\exp[-V_j])$.

The ODEs were integrated using a fourth-order Runge-Kutta
method.  The free parameters were scanned across
the following ranges:  for applied current $I_i$, 0-10 $\mu$A/cm$^2$; 
for $g_s$, the maximal synaptic conductance per cell,
0-2~mS/cm$^2$; for the synaptic decay time constant $\tau_s$,
5-50~ms.

\subsection{Calculation of coherence}

\begin{figure}[p]
\centerline{\epsfig{figure=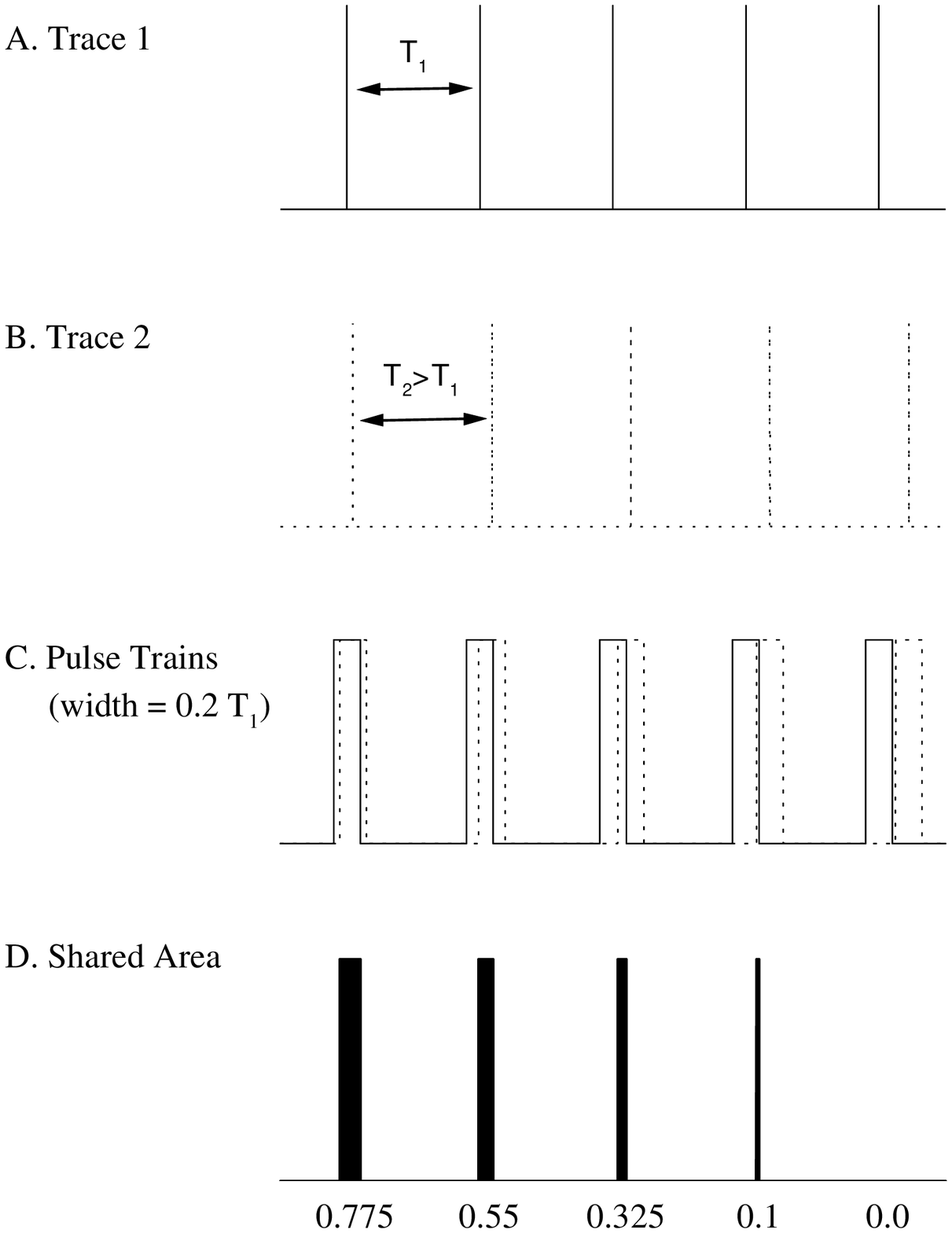, height=6.in,bbllx=12pt,bblly=12pt,bburx=599pt,bbury=780pt, clip=}}
\caption{An example of the coherence measure used in this work.
Panels A and B show idealized periodic spike traces with periods
$T_1$ and $T_2~>~T_1$.  
Panel C shows the pulse trains compared in the algorithm.
The solid line corresponds to Trace 1 and the dotted line to Trace 2.
Each pulse has unit height, width =~$0.2~T_1$, and is centered at 
the appropriate spike peak.
Panel D shows the shared area of the two pulse trains in graphical and
numerical form.}
\label{coho_ex_fig}
\end{figure}

As a measure of coherence between pairs of
neurons, we generated trains of square
pulses from the time domain responses of
each of the cells (Fig. \ref{coho_ex_fig}).
Each pulse, of height
unity, was centered at the time of
a spike peak (resolution = 0.1 ms); 
the width of the pulse was
20\% of the mean firing period of the
faster cell in the pair (0.2~$T_1$ in Fig. \ref{coho_ex_fig}). 
We then took the
cross-correlation at zero time lag of these pulse trains.
This is equivalent to calculating the shared area of the 
unit-height pulses, as shown in Fig. \ref{coho_ex_fig}D.
We took coherence as the sum of these shared areas, 
divided by the square root of the product of the summed areas of each
individual pulse train.
For the example shown in Fig. \ref{coho_ex_fig}, our algorithm gives coherence
of 0.35.

Our approach differs from the algorithm used
by Wang and Buzs\'aki~(1996), in which 
trains of unit-height pulses are correlated for
a bin width equal to or greater than
the neuronal time scale. 
The difference between the two algorithms can be appreciated
by considering the contribution made to the coherence 
measure by two spikes (in two separate neurons)
occurring with time difference $t_d$.
The Wang and Buzs\'aki (1996) algorithm would see these as
perfectly coherent if the spikes are in the same time bin
and incoherent if they are not.
The answer depends on where the bin edges fall, with 
probability of a coherence ``hit'' falling to zero when
the bin width is less than $t_d$.  In their algorithm
coherence is a function of the bin width, and averaging
across the population of cells ameliorates effects due to
the placement of bin edges.
In our algorithm, the two spikes make a
contribution to coherence that is continuously distributed
between 0 ($t_d >$ 20\% of firing period) and 1 ($t_d = 0$).
Although both algorithms give results that depend on 
the percentage of the firing period considered significant,
our measure allows us to examine coherence in small networks
with less discretization error.
This change is important here specifically because we
analyze small networks that phase-lock with a short but measurable
phase difference. 

We mapped coherence vs. $I_i$, $g_s$, and $\tau_s$ for networks 
of 2, 10, and 100 cells with all-to-all inhibitory coupling.
In networks with $N = 2$, coherence is plotted in the maps.
In larger networks, the plots show the average of the coherence
measure taken for all pairs of neurons.

\section{Results}

\subsection{Single self-inhibited neuron}

We first consider the firing characteristics of a single
self-inhibited neuron or, equivalently, a network of identical,
synchronized, mutually inhibitory neurons.
These simulations 
validate predictions from analytic work on simpler models~\cite{cc} 
and determine the ranges of
the phasic and tonic regimes in parameter space.
Firing frequency of the single neuron was tracked over the
parameter space of $I_i$, $g_s$, and $\tau_s$.  
Figure \ref{1cell_fig}A shows sample time-domain traces
for three values of $I_i$ (0.4, 1.6 and 9.0 $\mu$A/cm$^2$).
Like mammalian interneurons, the modeled 
system of differential equations produces
action potentials at rates up to 250 Hz.
Figure \ref{1cell_fig}B shows discharge frequency as a
function of $I_i$, for several values of 
$g_s$.  
For large values of $g_s$ (lower traces), this curve is roughly linear.
For smaller values (upper traces), discharge frequency rises along a
somewhat parabolic trajectory.
For negative values of $I_i$, the self-inhibited neuron can fire at
arbitrarily low frequencies (data not shown), indicative of a 
saddle-node bifurcation and synchrony through slow inhibition~\cite{bard96}.
In Fig.~\ref{1cell_fig}C
we show discharge frequency versus $\tau_s$ for several values of $I_i$,
with $g_s$ fixed.
The dependence of the frequency on $\tau_s$ for the lower two traces
is similar to what was
observed in the full network and {\it in vitro} by Whittington {\it et
al.}~(1995).  
The phasic and tonic regimes are clearly illustrated in
Fig.~\ref{1cell_fig}D, 
in which the ratio $\tau_s/T$ is plotted versus $\tau_s$ for
various values of $I_i$.  For large $I_i$ (top traces),
$\tau_s/T$ is large and linearly related to $\tau_s$,
indicative of the tonic regime.
In contrast, for small $I_i$ (bottom trace),
$\tau_s/T$ is small and depends 
only weakly on $\tau_s$, indicative of the phasic regime. 
For our model and level of heterogeneity, 
parameter sets that give $\tau_s/T < 1$ are in the
phasic regime; sets that give $\tau_s/T > 2$ are in the tonic regime.

Presence in either the phasic or tonic regime is dependent
on parameters other than $I_i$.
Generally, the tonic regime
is characterized by strong applied current and a relatively weak
synapse so 
that the firing period is much faster than the synaptic decay time.
The phasic regime occurs when either the applied current is weak and/or the
synapse is strong so that the firing period is locked to the decay
time.

\subsection{Two cell network}

We simulated networks of two 
mutually inhibitory cells with self-inhibition.  
We include
self-inhibition because it better mimics the behavior of a large
network.
In these and all other network simulations, mutual and 
self-inhibition are of equal weight.
In networks of two
interneurons with identical properties but different initial
conditions, the cells
quickly synchronize (phase-lock with zero phase difference) over the
entire examined range of $I_i$, $g_s$, and $\tau_s$ (data not shown).
Slow-firing cells tend to synchronize more quickly than
fast-firing cells, but the exact delay before synchronization
depends on initial conditions and was not examined systematically.
Anti-synchrony is not stable in the parameter regime we considered,
but could be with very small values of $\tau_s$.

\begin{figure}[p]
\centerline{\epsfig{figure=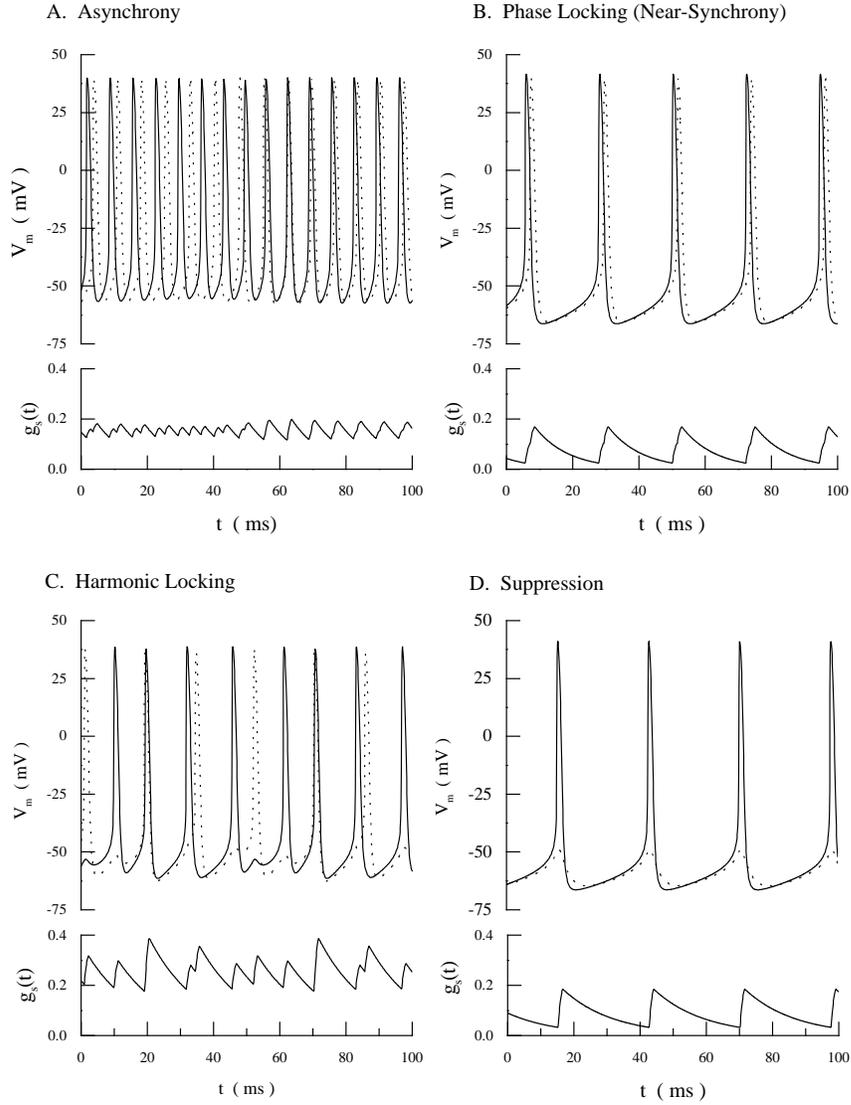,height=6.5in,bbllx=12pt,bblly=12pt,bburx=599pt,bbury=780pt, clip=}}
\caption{Plots of membrane potential ($V_m$) vs. time for two
heterogeneous neurons
at four different points in ($I_i,g_s,\tau_s$)-space.
In all cases, the solid (dotted) line is the more (less) 
excitable cell.
Also plotted in each panel is $g_s(t)$, the time-varying synaptic conductance
(in mS/cm$^2$) received by each of the two cells.
{\it A.}  Asynchrony with $I_1$ = 9.0, $I_2$ = 9.9 $\mu$A/cm$^2$;
$g_s$ = 0.25 mS/cm$^2$; $\tau_s$ = 10 ms.
{\it B.}  Near-synchrony with $I_1$ = 1.6, $I_2$ = 1.78 $\mu$A/cm$^2$;
$g_s$ = 0.25 mS/cm$^2$; $\tau_s$ = 10 ms.
{\it C.}  Harmonic locking with $I_1$ = 9.0, $I_2$ = 9.9 $\mu$A/cm$^2$;
$g_s$ = 0.5 mS/cm$^2$; $\tau_s$ = 10 ms.
{\it D.}  Suppression with $I_1$ = 1.6, $I_2$ = 1.78 $\mu$A/cm$^2$;
$g_s$ = 0.5 mS/cm$^2$; $\tau_s$ = 10 ms.
}
\label{domains_fig}
\end{figure}

When the input $I_i$ to each neuron is made mildly
heterogeneous (intrinsic spike rates $<$ 5\% different), 
a more complex picture emerges. 
Under the conditions of mild heterogeneity modeled here,
but not necessarily under conditions of greater heterogeneity~\cite{golomb93},
the behavior of the two-cell network falls into one of four
qualitative states, as exemplified by the traces of
membrane potential and inhibitory conductance vs. time in 
Fig.~\ref{domains_fig}.
For small $g_{s}$, large $I_i$, and large $\tau_s$ -- conditions
associated with the tonic regime -- the phasic component of synaptic
inhibition received by each cell is small (Fig. \ref{domains_fig}A).
The neurons influence each other's firing frequencies, but
firing times are independent.
We refer to this phase-dispersed state as the {\it asynchronous
state}.
As the phasic component of inhibition is increased, the phasic
regime is approached.
Within the phasic regime lie three qualitative states.
For appropriate choices of the level of inhibition, 
the two-cell network enters a phase-locked state with a non-zero
phase difference (Fig.~\ref{domains_fig}B).  We will continue to use the term
synchrony to refer to this near-synchronous regime. 
For this model, 
heterogeneity of some sort (in this case, heterogeneity of 
intrinsic firing frequencies) is a necessary and sufficient
condition for near, as opposed to pure, synchrony~\cite{cc}.
The size of the phase difference depends on the parameters chosen.
With further increases in the level of inhibition,
the faster cell begins to suppress its slower partner, leading
to what we term {\it harmonic locking} (Fig.~\ref{domains_fig}C).
In this example, cells
fire in a 4:3 ratio, and exert temporally complex effects
on each other during the course of one cycle ($\sim$ 50 ms).
Finally, with enough inhibition, the faster
neuron inhibits its slower counterpart totally, in what we term
{\it suppression} (Fig.~\ref{domains_fig}D).
In suppression, the sub-threshold dynamics of
membrane potential in the suppressed cell 
are exactly phase locked to those of the faster cell. 
This exact relationship holds because our simulations do not
include a synaptic delay term.

Without self-inhibition, this harmonic-locking regime is very small
and not seen in the analogous parameter space (data not shown). 
Our heuristic explanation for this difference is as follows.
Without self-inhibition, once the slower neuron is suppressed, 
the instantaneous preferred frequencies of the two cells diverge.
The faster cell is uninhibited and, by firing faster, adds more
inhibition to the slower cell, making it more difficult for the slower
cell to escape.
With self-inhibition, each of the cells in the two-cell network receives
an identical synaptic signal, effectively making the two cells more
homogeneous.
The added homogeneity increases the size of the region in which
harmonic locking occurs at relatively small locking ratios.

In order to observe network behavior over a large parameter range,
we used the relatively simple measure of firing coherence (see Methods).
A given level of coherence does not uniquely determine the qualitative
behavior of the network (asynchronous, synchronous, harmonic,
or suppressed).
However, the structures of coherence maps
are stereotyped, and coherence maps can be correlated to the four
qualitative network states.

Figures~\ref{coho_fig}A-B show three dimensional plots of coherence in a 
two-cell network, plotted versus $\tau_s$
and $g_s$ for low ($I_1=1.6$ and $I_2=1.78 \mu$A/cm$^2$)
and high ($I_1=9$ and $I_2=9.9 \mu$A/cm$^2$) applied currents.  
(The gray scale, which does not relate to coherence, is discussed
below.) 
Even though the differences in intrinsic (uncoupled) firing
frequencies for 
the two cells are small ($<5\%$ in each case), 
coherence is high and smoothly varying,
corresponding to  synchrony,
only over a small region of parameter space. 
The extent of the synchronous region increases as $\tau_s$ decreases.  
Increasing the heterogeneity reduces the size of the synchronous region.  
For differences
greater than a few percent in the intrinsic (uncoupled) frequencies, the
synchronous region was dramatically reduced in size (data not shown).

For a given $\tau_s$, synchrony is
broken in two distinct ways if $g_s$ is either too small or too large.  
For large $I_i$, large $\tau_s$, and (especially) small $g_s$, 
the phasic coupling between the two cells is weak and they fire
asynchronously (i.e., with dispersed phase).  
In this state, which is particularly large on the left side of 
Fig.~\ref{coho_fig}B, 
coherence has a value of about 0.2, corresponding to the expected value
of our coherence measure  with ``memory'' equal to 
20\% of the spiking period.
For large $g_s$, high levels of coherence are lost when the faster 
cell begins to suppress the slower cell, resulting in harmonic
spiking.
The particular pattern of harmonic spiking can change
dramatically with small changes in parameters, resulting in the
jagged coherence regions seen in Figs.~\ref{coho_fig}A-B.  Again, the harmonic
region is particularly noticeable with large $I_i$, as in Fig.~\ref{coho_fig}B.

Eventually, with large enough $g_s$, the full suppression 
state can take hold, and coherence plummets to give a very flat
region of coherence at a value of 0.
This state, favored by large $g_s$ and large $\tau_s$, 
occupies a large region on the right side of Fig.~\ref{coho_fig}A.

We argued in the discussion of Fig.~\ref{domains_fig} that the network's
presence in the asynchronous state is associated with the tonic
regime, and that the transition from asynchrony to 
locking is associated with the transition from the tonic
regime to the phasic regime.
To demonstrate this effect, 
we have gray-scale-coded the coherence maps
of Fig.~\ref{coho_fig} according to the value of $\tau_s/T$ obtained from
single, self-inhibited cells with the same values of
$\tau_s$ and total inhibition $g_s$ and $I_i$ taken as the average
of the range seen in the heterogeneous population.

The single-cell value of $\tau_s/T$ is useful as an indicator
of the qualitative state of all the cells in the network because
all the cells that are not suppressed fire at similar frequencies.
This result is demonstrated by 
Fig. \ref{compare_n}, which shows plots of $\tau_s/T$ for four conditions:
the N = 1 case (solid lines), the N = 2 case with differences
in intrinsic rates of around 4\% (dashed lines) and 2\%
(dashed-and-dotted lines); and the N = 10 case with maximal
heterogeneity of around 4\% (dotted lines).
In all cases with more than one cell, a pair of traces corresponding
to the fastest and slowest cells of the simulations are shown.
In all cases, the traces follow similar trajectories
until the slowest cell is suppressed 
(indicated by an abrupt end of the lower branchbefore
the rightmost point is reached).
This similarity in $\tau_s/T$ (and hence $T$) for all unsuppressed cells
is seen in both the phasic (Fig. \ref{compare_n}C) and tonic 
(Fig. \ref{compare_n}D, right side) regimes.

Returning to Fig.~\ref{coho_fig}, 
the value of $\tau_s/T$ as a predictor of transitions in qualitative
state and hence coherence implies that 
we should see transitions from asynchrony 
when $\tau_s/T$ drops below $\sim 2$.
As Figs.~\ref{coho_fig}A-B show, this approximate relationship does hold.
Furthermore, factors that change $\tau_s/T$ (e.g., changing $I_i$;
cf. Figs. \ref{domains_fig}A and \ref{domains_fig}B) 
have predictable effects on the extent of
the asynchronous state in ($\tau_s,g_s$)-space.

Figures \ref{coho_fig}C-D show similar results with less heterogeneity
($I_1 = 1.64, I_2 = 1.74 \mu$A/cm$^2$ for panel C; 
$I_1 = 9.2, I_2 = 9.7 \mu$A/cm$^2$ for panel D;
these values approximate the mean $\pm$ one standard deviation
for uniform distributions with limits as in Figs.~\ref{coho_fig}A-B). 
In these cases, the same qualitative coherence map is evident, with
a somewhat larger region of coherence.
The qualitative coherence regions correspond to the same qualitative
states from Fig.~\ref{domains_fig}.

\subsection{Large networks}

We also simulated all-to-all connected networks of 10 and 100
heterogeneous 
inhibitory neurons and found qualitatively similar results.  
Figures~\ref{coho_fig}E-F
show the coherence plots over the same parameter space as 
Figs.~\ref{coho_fig}A-B
for a network of 10 heterogeneous cells. 
The level of inhibition per synapse, $g_s/N$, scales with
$N$ to keep the level of inhibition per postsynaptic cell,
$g_s$, constant. 
For the ten-cell case, applied current $I_i$ is uniformly distributed
through the same ranges as in panels A-B ([1.6,~1.78] for panel
E; [9.0~9.9] for panel F).
Again, there are four qualitative states: an
asynchronous state for small $g_s$, more prevalent with higher $I_i$;
a near synchronous state; a harmonic state; and a suppressed
state.
For the 10 cell network, the transition to suppression is smoother
than in the two-cell case.  
Cells fall out of the rhythm to suppression one at a time,
leading to a relatively smooth drop in coherence.
At the highest values of $g_s$, coherence has not yet dropped
to zero because some cells are still able to synchronize with
the fastest neuron of the network.
In the harmonic state, examination of time-domain traces 
(data not shown) reveals harmonic patterns, with a cluster of
cells in  synchrony while the slower cells drop in and
out of the population rhythm.
The coherent region for the ten-cell network is larger than in
Figs.~\ref{coho_fig}A-B.
Applied currents (and hence intrinsic frequencies) of the two 
neurons in panels A-B are at the limits of the range of 
applied currents in the ten-cell network, making the 
effective level of heterogeneity smaller in the ten-cell case.
The close agreement between panels C-D and E-F supports this
contention.

We also performed a limited number of simulations of a 
100-cell network with the same architecture, at parameter values
representing orthogonal slices through the 3-dimensional
coherence maps.
Results from these simulations are shown in Fig.~\ref{slices_fig},
along with slices from the coherence maps of Fig.~\ref{coho_fig}.
In Figs.~\ref{slices_fig}A-B, coherence is plotted vs. $g_s$ for a 
fixed value of $\tau_s$ = 15~ms and at two levels of applied
current.
In Figs.~\ref{slices_fig}C-D, coherence is plotted vs. $\tau_s$ for a fixed value
of $g_s$ = 0.5~mS/cm$^2$.
Results from the 100-cell (N~=~100) and 10-cell (N~=~10) 
cases are quite similar,
at both low (panels A, C) and high (panels B, D) levels of
applied current.
These results support the argument that the qualitative behavior
of the network does not change with N, and thus that predictions
based on single-cell analysis and simulations are applicable to
moderately heterogeneous networks of arbitrary size.
Results are shown for both levels of heterogeneity in 2-cell networks.
The dashed lines (N~=~2), which are slices through the coherence maps
of 
Fig.~\ref{coho_fig}A-B, have lower coherences that reflect the relatively large
amounts of heterogeneity in these cases.
The dashed-and-dotted lines (N~=~2*) show coherence values for
slices through Figs.~\ref{coho_fig}C-D, with closer intrinsic frequencies
chosen to approximate the standard deviations of the appropriate
uniform distributions.
These slices more nearly match the 10- and 100-cell cases.

Results from Figs. \ref{compare_n} and 6 also demonstrate the close relationship
between the ratio $\tau_s/T$ and coherence (as well as underlying
qualitative states).
Values of $\tau_s/T < 2$ from Fig. \ref{compare_n} 
are almost invariably associated
with one of the locked states.
Values of $\tau_s/T > 2$, on the other hand, give rise to the 
asynchronous state, associated in Fig.~\ref{slices_fig} with regions of flat
coherence at a value of 0.2 (e.g., the leftmost portion of 
Fig.~\ref{slices_fig}B
and the rightmost portion of Fig.~\ref{slices_fig}D).

\section{Discussion}

We show that the behavior of the firing frequency of a single
self-inhibited cell can give insight into the network
frequency and coherence.  
In particular, the ratio of the synaptic decay constant $\tau_s$
to the neuronal firing period $T$ has rough predictive value
in determining whether a mildly heterogeneous network is
synchronous or asynchronous.
This predictive value only holds with mild heterogeneity,
however; greater heterogeneity leads to a mixture of qualitative
states~\cite{golomb93} which invalidates our analyses.

We also emphasize the importance of even mild 
heterogeneity in affecting network dynamics.  
Previously, it had been argued   
that slowly decaying inhibition generally had a synchronizing
influence~\cite{gerstner96b,terman96,vrees94}.  
However, for mildly heterogeneous cells, 
the relation of the frequency (or period) 
to the synaptic decay time must also be
considered.  
For homogeneous cells, the synaptic coupling is only
required to align the phases in order to obtain synchrony.  
For mildly heterogeneous cells, 
the coupling must both align the phases and
entrain the frequencies.  
The latter is more difficult for the network to achieve.
It occurs only when the inhibition is strong enough so that firing
period is 
dominated by the decay time.  However, if the inhibition
is too strong then the slower cells will never fire.
Thus, there are
two ways to destroy full network synchrony.  
The first is through
effective de-coupling where the cells tend to fire asynchronously.
The second is through suppression, in which the neurons with
higher intrinsic rates fire in near-synchrony and
keep their slower counterparts from firing.  
Between synchrony and
suppression harmonic locking is also possible.  This occurs when
the suppression of the slower cell is temporary but lasts longer
than the period of the faster cell.
We should note that anti-synchrony, not seen in the
parameter regimes presented here,
can become stable with very fast 
synapses (i.e., $\tau_s/T 
<< 1$)~\cite{friesen94,perkel74,skinner94,vrees94,wang92}.

\begin{figure}[p]
\centerline{\epsfig{figure=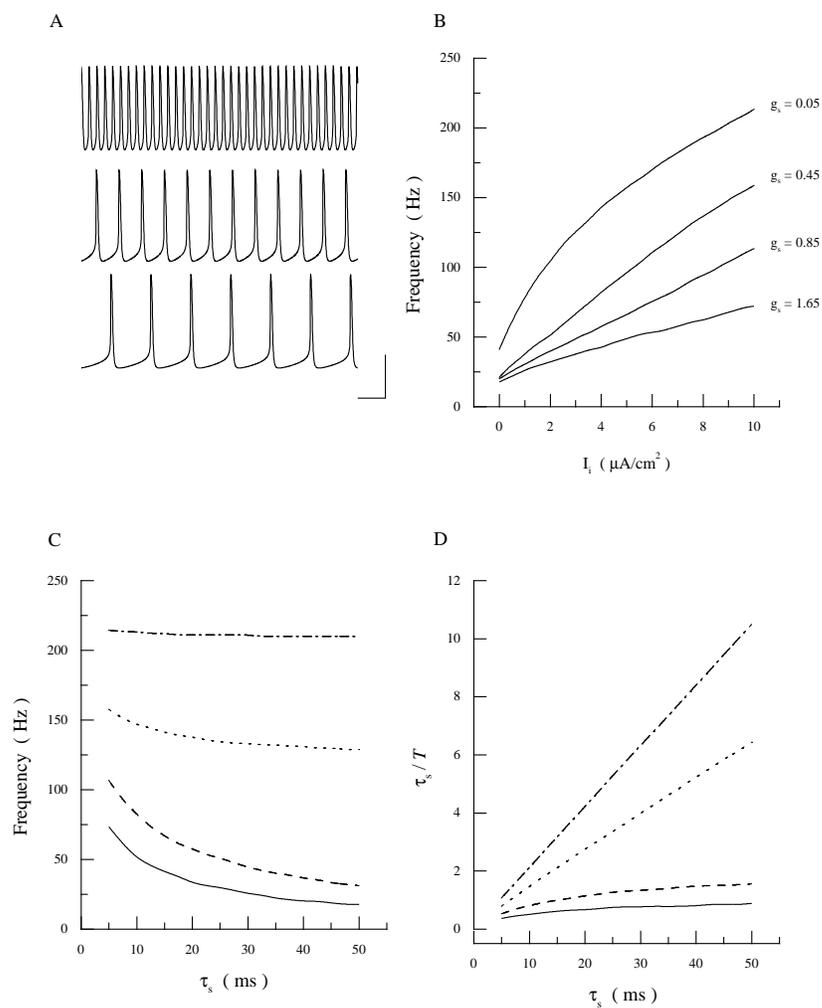, height=6.in,bbllx=12pt,bblly=12pt,bburx=599pt,bbury=780pt, clip=}}
\caption{Behavior of the single, self-inhibited neuron.
{\it A.}~~Time-domain responses of the self-inhibited neuron
($g_s$ = 0.25 mS/cm$^2$, $\tau_s$ = 10 ms) for three values
if $I_i$ (from bottom to top: 0.4, 1.6, and 9.0 $\mu$A/cm$^2$).
Horizontal scale bar: 20 ms.  Vertical scale bar: 50 mV.
{\it B.}~~Neuronal discharge frequency vs. applied current $I_i$
for several values of $g_s$ (from top to bottom: 0.05, 0.45,
0.85, 1.65 mS/cm$^2$).  $\tau_s$ = 10 ms.
{\it C.}~~Firing frequency vs. $\tau_s$.
From bottom to top, $(g_s, I_i)$ = (0.45,2.0) (solid line), 
(0.45,4.0) (dashed line), 
(0.2,6.0) (dotted line), and (0.05,10.0) (dashed-and-dotted line).  
Conductances have units of mS/cm$^2$.  
Currents have units of $\mu$A/cm$^2$.
{\it D.}~~The ratio of the synaptic decay time constant ($\tau_s$)
to the neuronal discharge period ($T$), plotted vs. $\tau_s$.
Different line types represent the same values
of $g_s$ and $I_i$, in the same order, as in panel {\it C}.
}
\label{1cell_fig}
\end{figure}

For even mildly heterogeneous cells, synchrony in which all inhibitory
cells participate is possible only over a small
region of parameter space 
that decreases as the heterogeneity is increased.  The region where
synchrony occurs in a large network of known (mild) heterogeneity and 
connectivity can be approximated from a two-cell network.  
The frequency of firing and conditions allowing synchrony 
can be estimated analytically from a reduced model neuron with
self-inhibition~\cite{cc}
As in large networks~\cite{traub96}, the frequency in single cells 
depends on the applied
current, the synaptic strength and the synaptic decay time.  
In the synchronous region, the firing period depends linearly on the
decay time and logarithmically on the other parameters so that
frequency will depend directly on the decay rate.  However, 
the contribution from the logarithmic factor can be fairly
large and thus must be calculated explicitly.  This can be estimated
analytically from the reduced model~\cite{cc}, or from simple
simulations of a single, self-inhibiting cell (see Fig.~\ref{1cell_fig}).

\begin{figure}[p]
\centerline{\epsfig{figure=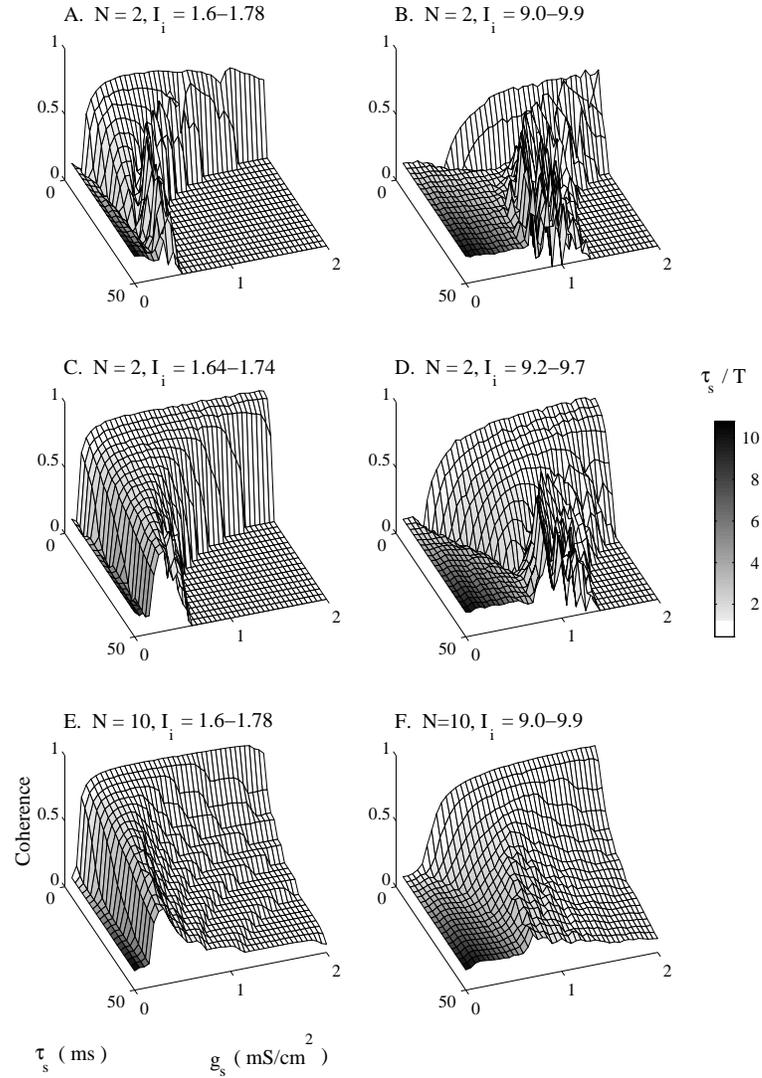, height=7in,bbllx=12pt,bblly=12pt,bburx=599pt,bbury=780pt, clip=}}
\caption{Coherence maps in ($\tau_s,g_s$)-space.
Top row: Coherence vs. $\tau_s$ and $g_s$ for the two-cell
network, with $I_1$ = 1.6 and $I_2$ = 1.78 $\mu$A/cm$^2$ ({\it A}),
$I_1$ = 9.0 and $I_2$ = 9.9 $\mu$A/cm$^2$ ({\it B}).
Middle row: Coherence vs. $\tau_s$ and $g_s$ for the two-cell
network, with $I_1$ = 1.64 and $I_2$ = 1.74 $\mu$A/cm$^2$ ({\it C}),
$I_1$ = 9.2 and $I_2$ = 9.7 $\mu$A/cm$^2$ ({\it D}).
Bottom row: Coherence maps for ten cells with $I_i$ uniformly
distributed in the intervals [1.6,~1.78] ({\it E}),
[9.0,~9.9] ({\it F}).
In all maps, the gray scale gives the ratio $\tau_s/T$ (see scale bar).
}
\label{coho_fig}
\end{figure}

\begin{figure}[p]
\centerline{\epsfig{figure=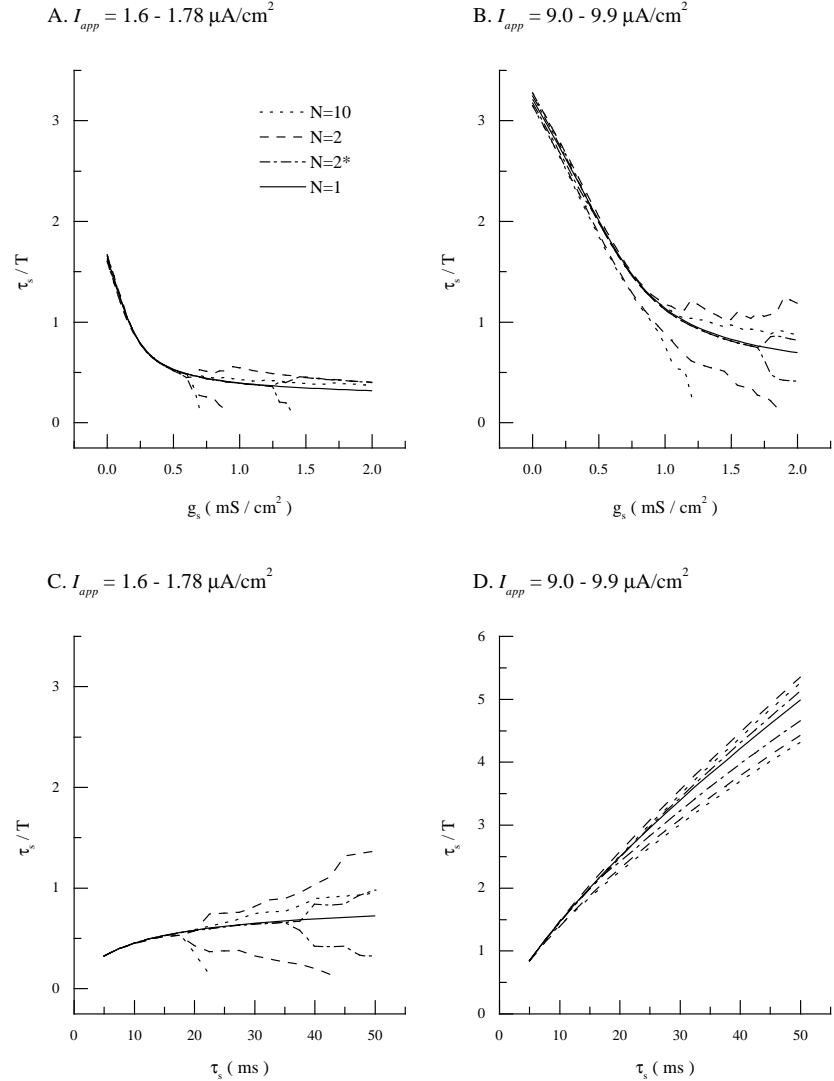,height=6.5in,bbllx=12pt,bblly=12pt,bburx=599pt,bbury=780pt, clip=}}
\caption{Plots of $\tau_s/T$ vs. $g_s$ ({\it A-B}) and $\tau_s$ ({\it C-D})
for networks of size N = 1, 2, and 10. 
For the N = 2 (dashed lines) and N = 10 (dotted lines) cases, 
values of $I_i$ were evenly distributed between the inclusive limits shown,
as in Figs.~\ref{coho_fig}A-B and \ref{coho_fig}E-F,
giving about 4\% maximum heterogeneity in intrinsic firing rates.
For the N = 2* case (dashed-and-dotted lines), $I_1$ and $I_2$
were set to the same values as in Fig.~\ref{coho_fig}C-D,
giving about 2\% heterogeneity.
For N = 1, $I_i$ is the center point of the interval.
For all cases with N $> 1$, two traces are shown, representing values
from the fastest and slowest neurons from the simulations.
Suppression of the slowest cell is represented by early terminations
of the curves.
}
\label{compare_n}
\end{figure}

\begin{figure}[p]
\centerline{\epsfig{figure=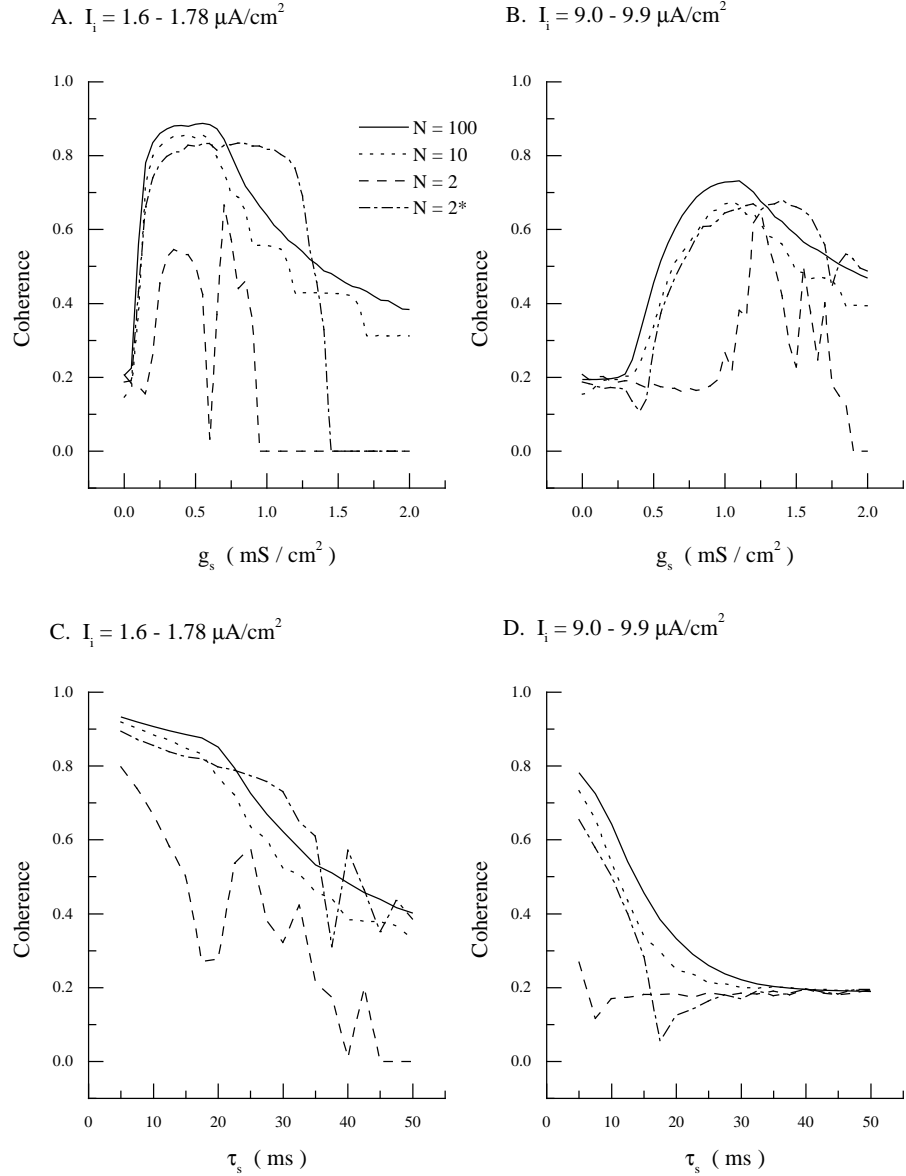, height=7in,bbllx=12pt,bblly=12pt,bburx=599pt,bbury=780pt, clip=}}
\caption{Two-dimensional slices through coherence maps.
Left column: coherence vs. $g_s$ at $\tau_s$ = 15 ms, 
for $I_i$ uniformly distributed 
in the ranges [1.6,~1.78] ({\it A}) and [9.0,~9.9] ({\it B}).
Shown are coherence for 100 cells (solid line), 10 cells (dotted line;
data from Fig.~\ref{coho_fig}E-F),
2 cells at the limits of the distribution of $I_i$ (dashed line;
data from Fig.~\ref{coho_fig}A-B),
and 2  cells at intermediate values of $I_i$ (dashed-and-dotted line;
data from Fig.~\ref{coho_fig}C-D). 
Right column: coherence vs. $\tau_s$ at $g_s$ = 0.5 mS/cm$^2$,
for low ({\it C}) and high ({\it D}) values of $I_i$ (specific
values as in {\it A} and {\it B}, respectively).  
Line types have the same meaning as in {\it A-B}.  
}
\label{slices_fig}
\end{figure}

The result that the value of $\tau_s/T$ from single-cell 
simulations has predictive value for the qualitative state and
coherence of a network of arbitrary size is intriguing and potentially
useful, because
it points the way to determining the qualitative and quantitative
behavior of a neuronal network based on simple behavior that
can be studied numerically or even analytically.
However, the predictive capabilities of this index should not
be overestimated.
A careful examination of Fig.~\ref{coho_fig} shows that 
the mapping between $\tau_s/T$ and asynchrony is not
precise.  The value of $\tau_s/T$ at which the transition will
occur is dependent on many factors, including the level of
heterogeneity and, in all likelihood, the level and form 
of connectivity in the network.
The value of $\tau_s/T$ alone is not sufficient to determine
the point of transition from synchrony to harmonic locking
and suppression, even in a model of known heterogeneity and
architecture.
Making this determination requires knowledge of $I_i$ and $g_s$
in addition to $\tau_s/T$~\cite{cc}.

Studies of the 2-cell network were successful
in elucidating the qualitative states of the larger circuit,
though the exact form of transitions 
from asynchrony to synchrony and synchrony to suppression is
different in detail for our simulations of the 2-cell and N-cell cases.
In general, the behavior of the 2-cell network matches that of the
N-cell circuit better in the asynchronous state, associated with
the tonic regime, than in the harmonic and suppression states,
associated with the phasic regime.
This result is expected from our theoretical framework since
the tonic regime is defined as the regime in which
only the tonic level of inhibition is important.  Since we
normalized the synaptic strength by N, the net amount of inhibition is
independent of the network size.
Thus, we take this result as additional evidence that our
hypothesized mechanisms of loss of coherence are correct.

Our numerical results are similar to those of 
Wang and Buzs\'aki~(1996), but our explanations differ considerably.
In heterogeneous networks, they also saw a decline in coherence with
both low and high firing rates.  
They attributed the decline in synchrony for low rates to two factors.
First, they point out that cells are more sensitive at low firing rates 
than at higher rates to changes in applied
current, a source of heterogeneity in both studies.
This point is correct, but in our work we controlled for this
factor, using smaller percent differences in small currents than
in large currents to achieve similar percent differences in
intrinsic firing rates, and we still saw a drop-off in coherence
at low rates.
Second, Wang and Buzs\'aki (1996) cite what they call a 
``dynamical'' effect, in which inhibition is fast enough to destabilize
the synchronous state.
Previous work~\cite{vrees94,wang92} shows that the outcome of 
such dynamical effects for
{\em homogeneous} networks is anti-synchrony.
In our parameter regime, the loss of coherence in 
{\em heterogeneous} networks at low firing rates 
(i.e., with $\tau_s/T$ small) is associated with the phasic regime
and is due to suppression of firing in slower cells.
Wang and Buzs\'aki~(1996) make the phenomenological argument
that the loss of synchrony
at high firing rates is related to a need for greater density of
synaptic connectivity. 
We considered all-to-all connectivity and found that
loss of coherence 
associated with high firing rates (tonic regime)
is caused by a loss of too much of the phasic component of inhibition.
Furthermore, we argue that one can approximate the parameters for which 
this loss of coherence occurs by analyzing the single, self-inhibitory
cell. 
It should be possible to generalize these results and arguments to
the case with less than all-to-all coupling.

It has been suggested that the selection of the network frequency
{\it in vivo} is determined by the tonic
excitation and the parameters regulating the synaptic
coupling~\cite{traub96}.  Our results support this hypothesis.
However, we have demonstrated that with heterogeneous cells,
synchrony may not be possible at all frequencies.  
In particular, a network of this kind seems unlikely to support
synchronous firing at 200 Hz, a frequency that seems too fast
to be synchronized by GABA$_A$ receptors with $\tau_s \sim 15$ ms
(and $\tau_s/T \sim 3$).
Our framework implies that this result, which has been seen in
simulations 
before~\cite{wang96}, holds in general for heterogeneous cells in the 
tonic regime.

Our results emphasize the difficulty of generating synchronous 
oscillations in
interneuronal networks over a large range of frequencies,
such as in the transition from the gamma/theta mode to the
sharp wave/fast ripples mode.
At gamma frequencies, the factor $\tau_s/T$ should be less than 1
with typical values of $\tau_s$.
Thus, full synchrony at gamma frequencies is possible but requires
careful regulation of the system to prevent suppressive effects.
The question of whether or not the suppression we see is incompatible 
with physiological data cannot be answered, because it is extremely
difficult to estimate the number of interneurons participating in the
rhythm.
We believe that this issue can be explored, and our model tested, by
examining the power of the gamma field potential in a brain slice as
$\tau_s$ is modified by pentobarbital.
Our model predicts that the power in this signal should decrease as 
$\tau_s$ rises
and suppression becomes more evident.
A negative result in these experiments would indicate that our model is
missing a fundamental element.
One such element is intrinsic or synaptic noise, which can act to release
neurons from suppression (White, unpublished observations).

The more difficult goal for our model to achieve is that of
firing synchronously at ripple (200 Hz) frequencies, as has been
reported in the behaving animal~\cite{ylinen95}.
One or more of several conceivable explanations may underlie 
this apparent robustness in hippocampal function at high frequencies.  
First, it is possible, but unlikely, that heterogeneity in 
the intrinsic firing frequencies of interneurons is very low
($ < 4$\%).
Second, the operant value of $\tau_s$ may be lower than we
believe; a value of 5 ms would conceivably allow synchrony
at 200 Hz with levels of heterogeneity of around 5\%.
Third, each interneuron may fire not at 200 Hz, but rather
at a lower frequency of, say, 100 Hz, during sharp waves.
Under this explanation, the 200-Hz ripple would be generated
by clusters of two or more populations of neurons spiking
independently.
Finally, some factor(s) not considered here may enhance synchrony
at high frequencies.
Gap junction-mediated electrical coupling among interneurons, 
for which some evidence exists in the hippocampal region CA1~\cite{katsu},
is perhaps the most likely such factor~\cite{traub95}.

\acknowledgements
We thank M. Camperi for assistance in writing code, 
and S. Epstein, O. Jensen, C. Linster, and F. Nadim for helpful
discussions.  
B. Ermentrout, J. Rinzel, and R. Traub provided valuable feedback
on earlier versions of the manuscript.
This work was supported by grants from 
the National Science Foundation (DMS-9631755 to C.C., N.K. and J.W.), 
the National Institutes of Health (MH47150 to N.K.; 
1R29NS34425 to J.W.), 
and The Whitaker Foundation (to J.W.)

\end{document}